\documentclass[useAMS,usenatbib]{mn2e}

\usepackage{psfig}
\usepackage[dvips]{graphicx}
\usepackage{lscape}

\def\lesssim{\mathrel{\hbox{\rlap{\hbox{\lower4pt\hbox{$\sim$}}}\hbox{$<$}}}}
\def\gtrsim{\mathrel{\hbox{\rlap{\hbox{\lower4pt\hbox{$\sim$}}}\hbox{$>$}}}}

\title[The evolutionary sequence of subm-mm galaxies]{The evolutionary sequence of sub-mm galaxies: \\
from diffuse discs to massive compact ellipticals?}
\author[Ricciardelli et al.]{E. Ricciardelli$^{1,2}$\thanks{E-mail: elenaricci@iac.es}, I. Trujillo$^{1,2}$, F. Buitrago$^{3}$ and C. J. Conselice$^{3}$\\
$^{1}$Instituto de Astrof\'isica de Canarias, V\'ia Lactea s/n, E-38200 La Laguna, Tenerife, Spain\\
$^{2}$Departamento de Astrof\'isica, Universidad de La Laguna, E-38205, Tenerife, Spain\\
$^{3}$School of Physics and Astronomy, University of Nottingham, NG7 2RD, UK}

\begin{document}
\date{Accepted 2010 March 15.  Received 2010 March 15; in original form 2009 October 15
}
\pagerange{\pageref{firstpage}--\pageref{lastpage}} \pubyear{2009}
\maketitle
\label{firstpage}
\begin{abstract}
The population of compact massive galaxies observed at $z > 1$ are hypothesised, both
observationally and in simulations, to
be merger remnants of gas-rich disc galaxies. To probe such a scenario 
we analyse a sample of 12 gas-rich and active star forming sub-mm galaxies (SMGs) at $1.8<z<3$. We
present a structural and size measurement analysis for all of these objects 
using very deep ACS and NICMOS imaging in the GOODS-North field. Our analysis reveals a heterogeneous mix of
morphologies and sizes. We find that four galaxies (33\%$\pm$17\%) show clear signs of mergers or
interactions, which we classify as early-stage mergers. The remaining galaxies are
divided into two categories: five of them (42\%$\pm$18\%) are diffuse and regular disc-like
objects, while three (25\%$\pm$14\%) are very compact, spheroidal systems. We argue that these
three categories  can be accommodated into an evolutionary sequence, showing the
transformation from isolated, gas-rich discs with typical sizes of 2-3 kpc, into compact ($\lesssim$ 1 kpc) 
galaxies through violent major merger events,  compatible with the scenario depicted by
theoretical models.  Our findings that some SMGs are already dense and compact
provides strong support to the idea that  SMGs are the precursors of the compact, massive galaxies found at slightly lower redshift. \\
\end{abstract}

\section{Introduction}\label{intro}

Observations have shown that massive spheroids at high redshift are remarkably 
smaller than their local counterparts 
(e.g. \citealt{Daddi:05, Tr:06a, Tr:07, Longhetti:07, Buitrago:08, Cimatti:08, Vand:08}). 
This population of galaxies must evolve into local massive ellipticals based on their
stellar masses, however it is 
not yet clear what the primary mechanism is which increases their sizes by a factor of four 
to match the local galaxy population. Although various effects (i.e. dry major mergers, 
observational biases, selection effects, see for instance \citealt{Valentinuzzi:09}) may 
explain the observed size evolution of these systems, a physically motivated favoured 
mechanism is the growth in size by later minor mergers with less dense 
galaxies \citep{Bournaud:07, Naab:09, Hopk:09a}.  This scenario has the
advantage in that it facilities the growth in galaxy sizes while permitting a
mild evolution in velocity dispersion as observed (e.g. \citealt{Cenarro:09, Cappellari:09}).

An additional problem  is understanding how these objects first formed. An emerging
picture (e.g., \citealt{Hopk:07, Cimatti:08, Hopk:09}) for  the formation of these compact 
systems predicts that massive, gas-rich galaxies at very high redshift become unstable following
a major merger event, triggering a short-lived starburst within $\sim
0.1 $Gyr. Theoretical models \citep{KS:06, Hopk:07} have shown that  the size of the
remnant strongly depends upon the degree of dissipation involved, being very small in the
case of strongly dissipative mergers. Since at high redshift ($z \gtrsim 2$) galaxies
are more gas-rich than they are today \citep{Erb:06}, the degree of dissipation is expected 
to be high, and the resulting remnant extremely compact, with sizes $\lesssim 1$ kpc. 

Given the great amount of gas involved in these star formation processes,  we expect the 
progenitors of massive compact galaxies to be undergoing a high amount of star
formation, and hence should  be detectable in the subm-mm \citep{Narayanan:09}.   To
tests this hypothesis we examine in this paper a sample of SMGs (sub-mm galaxies), which
have been imaged with HST in deep exposures, to probe their structural properties.  SMGs are 
among the most luminous ($L \simeq 10^{13} L_{\odot}$), and  rapidly star forming ($SFR \simeq
10^{3}M_{\odot}/yr$) galaxies in the high-redshift universe 
 \citep{Hughes:98, Eales:00, Smail:02, Coppin:06, Tacconi:06,Tacconi:08, Menendez:09} and are
believed to be the precursors of local early-type galaxies
\citep{Sw:06}.  As their local counterparts, the ultra-luminous infrared galaxies, the
origin for their high fluxes is thought to be a  strong starburst and/or AGN activity, 
likely triggered by a major merger \citep{SM:96, Murphy:96, Clements:96}. 
It is also plausible that many sub-mm galaxies are similar to the hyper-luminous 
infrared galaxies (HLIRGS, $L_{IR}>10^{13}L_{\odot}$, \citealt{RR:00}), containing a
similar amount of molecular gas, and an extreme star formation rate \citep{Farrah:02a, Farrah:02b}. 

Furthermore, the similarity between the stellar mass surface densities of SMGs
and the compact massive galaxies at lower redshifts \citep{Tacconi:06, Tacconi:08, Cimatti:08} 
makes sub-mm galaxies natural candidates for being the precursors of these compact galaxies.
This paper tests this  hypothesis and investigates whether the proposed theoretical scenario
for the formation of massive compact galaxies is compatible with present deep
imaging and star formation analysis of these sub-mm systems.   To explore this, we
probe the sizes of the SMGs in various phases, which we construct based on morphology
and structure, and test whether the star formation rates of these galaxies are in agreement with 
theoretical expectations. Second, we probe whether the
different observed morphologies of the SMGs can be fitted into the evolutionary
sequence proposed by current massive galaxies formation models. 

The paper is structured as follows. 
In Sect. \ref{data} we describe our sample. In Sect. \ref{anal} we explain the method adopted for our analysis. In  Sect. \ref{res} we present our results and discuss their implication in Sect. \ref{disc}.
Throughout, we assume the following cosmology: $H_0$=70 \mbox{km s$^{-1}$} \mbox{Mpc$^{-1}$}, 
$\Omega_m=0.3$ and $\Omega_{\Lambda}=0.7$ and we use AB magnitudes. 

\section{Data}\label{data}

\begin{figure}
\begin{center}
\includegraphics[width=0.75\columnwidth]{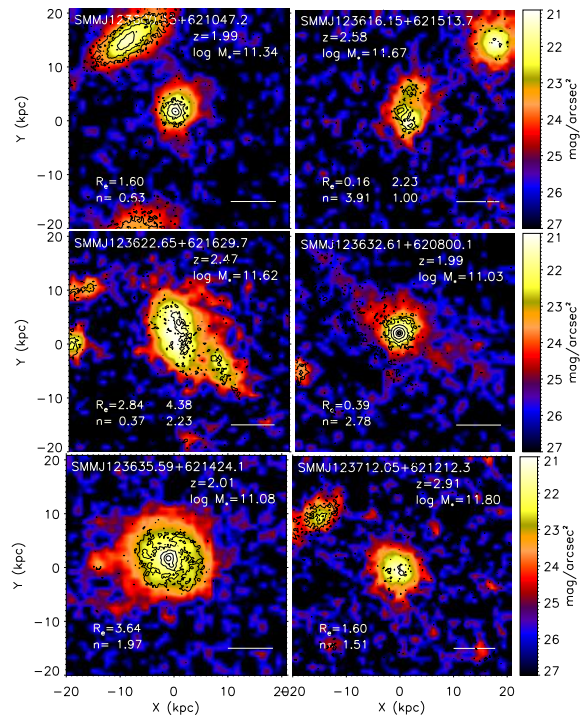}\\
\hspace{1in}
\end{center}
\caption{NICMOS images of our sample of SMGs with ACS contours overlapped. Contour levels of 24,23,22,21 and 20 mag/arcsec$^2$ are shown with increasing thickness. 
For each galaxy we indicate redshift, logarithmic stellar mass (in units of solar masses), effective radius (in kpc)  and S\'ersic index. The white line in the lower-right corner is 1 arcsec in size. }
\label{nic}
\end{figure}

Our target selection is based on the following criteria, needed to characterize the nature of our
high redshift sub-mm galaxies: (i)  very deep images (particularly with ACS on the Hubble Space
Telescope) to detect signatures of interactions;  (ii) systems within and similar to the redshift range where the massive 
compact galaxies have been detected (i.e. $2\lesssim z \lesssim 3$), ideally based on spectroscopic redshifts; 
(iii) our SMGs should be massive ($M>10^{11}M_{\odot}$) to ensure that these objects are indeed the progenitors
of the lower redshift massive galaxy population.

For the above reasons the number of SMGs that we can explore is limited. We adopt as our parent sample the 
SMGs from \citet{Mich:09}.  This paper contains spectroscopic redshifts (from \citealt{Chapman:05}),
multi-wavelength  photometric data-points collected from the literature, including Spitzer observations.
This allows for the determination of accurate SFRs and stellar masses for 76 massive SMGs (see
\citet{Mich:09} for these values and the details in how they were calculated).   The resulting range
of stellar mass is quite small: $10^{11}-10^{12} M_{\odot, }$. For these reasons this sample of
sub-mm galaxies are the progenitors of a fraction of local massive ellipticals, as well as the 
 compact galaxies at high redshift. Since we are interested in the high-redshift population, 
we restrict our analysis to the redshift range: $1.8<z<3$. The original stellar masses from this catalogue are
converted from a Salpeter Initial Mass Function (IMF) to a Kroupa IMF \citep{Kroupa:01}.

Our new analysis is based on deep archival HST ACS and NICMOS images of these sources, where we 
find images for 12 sources  in the GOODS North field.  We also investigated sources in other
fields (e.g.,  \citealt{Webb:03, Clements:04}), but these have too shallow observations 
to be useful for our purposes\footnote{These observations typically have integration 
times $<$ 7000s compared to the $>$ 27000s for the GOODS-N imaging}.

\begin{figure*}
\begin{center}
\includegraphics[width=1.6\columnwidth]{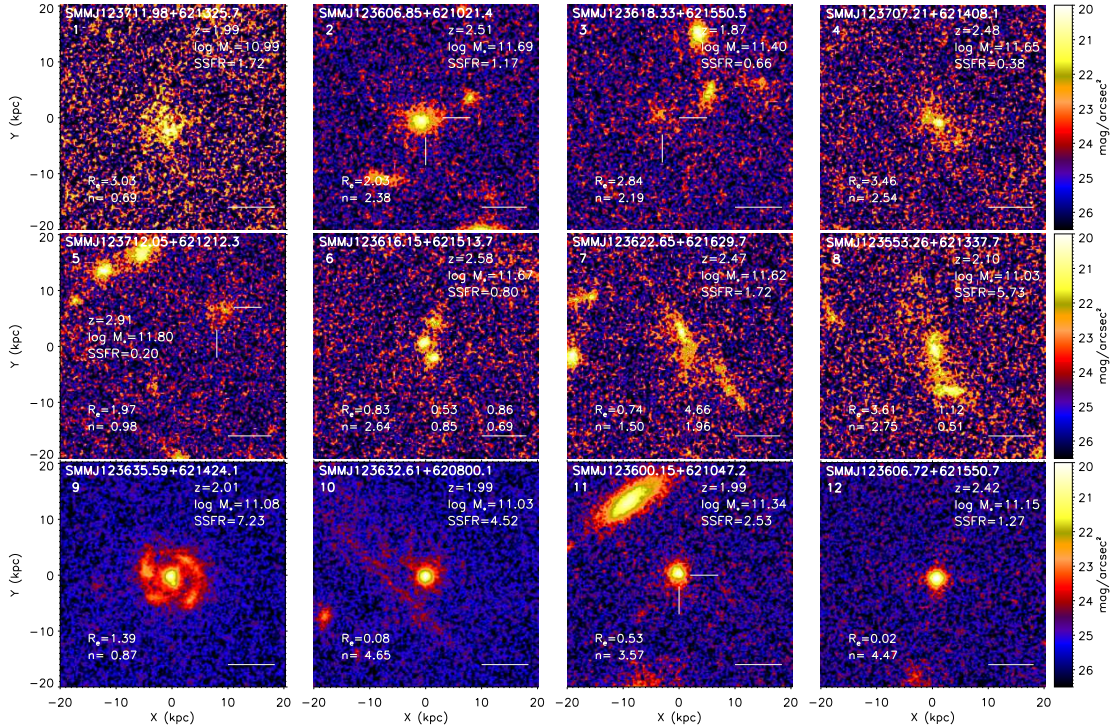}\\
\hspace{1in}
\end{center}
\caption{ACS images of our sample of 12 SMGs. The galaxies are ordered according to the evolutionary sequence described in the Sec. 4. The first five, from left to right and
top to bottom, are the disc-like objects, the next four are the ongoing mergers, and the final three are our classified compact systems. Each panel shows a box-size of 40x40 kpc. The white line in the lower-right corner shows the 1 arcsec scale-length. In the cases where the target object is not obvious it is indicated by two white bars. 
As in Figure \ref{nic} stellar masses are in units of solar masses and effective radii in units of kpc. Specific star formation rates (SSFR) are in \mbox{Gyr$^{-1}$}.  }
\label{acs}
\end{figure*}

Our final collection of images is as follows: we have ACS images in the F850LP filter (z-band)  for all 12 objects, and
NICMOS NIC-3 data from the GOODS NICMOS Survey (GNS; \citealt{Bluck:09}; Conselice
et al. 2010 in preparation) in the F160W filter (H-band) for 6 objects.  The z-band data, at 5
orbits depth, reaching a magnitude limit of z=27 mag (15$\sigma$), have been drizzled to a scale of 
0.03'' with a Point Spread Function (PSF) Full Width at Half-Maximum of 0.1''. The NICMOS images, at 3 orbits 
depth (limiting magnitude of H=26.8 mag, 5$\sigma$ ), were combined to
produce images with a pixel scale of 0.1'' and PSF size of 0.3''. Note that at $z \sim 2.5$
the ACS data trace the near-UV rest-frame, whereas the NICMOS imaging shows the B-band
rest-frame. The NICMOS and ACS images of the final sample are shown in Figure \ref{nic}
and \ref{acs}, respectively. 

All the objects in our sample show hints of AGN activity either from optical spectral features, 
or from X-ray and radio detections.  However, it is unlikely that these AGN dominate the bolometric 
emissivity in such objects. The reason for this is that all the SEDs are well-fit by pure star forming models,  
without the need of an AGN contribution \citep{Mich:09}. The mid-IR spectra of these sub-mm galaxies 
further confirm that they are starburst dominated without a bolometrically significant AGN \citep{Pope:08}.
Moreover, as shown by \citet{Alexander:05}, the X-ray to far-IR luminosity ratio is much lower than 
that found in QSOs, indicating that the contribution of the AGN to the total luminosity can not be higher than 10\%.
More recently, \citet{Laird:10} found that the X-ray emission in SMG is largely due to star formation activity and, even in the cases where the presence of AGN is confirmed, it is not the dominant contributor to the bolometric luminosity, except in rare cases.
Therefore we expect that the properties of our sub-mm selected galaxies are not biased by AGN activity.

\section{Analysis}\label{anal}

Morphological parameters and sizes are measured on our sample through the two-dimensional 
fitting code GALFIT \citep{Peng:02}. We model the light distribution of the sources with a
S\'ersic profile, deriving the  S\'ersic index $n$, the axis ratio $b/a$ and the
semi-major effective radius $r_e$  in arcsecs. We scale the radius to the physical
scale in kpc relative to the redshift of the source  and we circularise it
through $R_e=r_e \sqrt{b/a}$. The robustness of GALFIT in recovering sizes and structural parameters
using HST data was assessed in several previous papers (e.g. \citealt{Tr:06b, Tr:07,
Cimatti:08}), through the use of simulated galaxies, and is found to be robust
at the resolution and depth of our HST imaging.

To check the reliability of our measurements against changes in the PSF shape along the
images we use up to five different natural stars, as found in our fields, as PSFs.   We measure
the structural parameters for each objects five times, taking as our final measurement the
biweight estimator, and its confidence interval as the uncertainty in the measurement.
Most of the sources have an uncertainty of $\lesssim 20 \%$ in the measured size and 
$\lesssim 30 \%$ in the S\'ersic index.  Only two cases (SMMJ123606.72+621550.7 and
SMMJ123632.61+620800.1) contain large errors based on the ACS imaging analysis, 
as their  effective radii are close to the size of the PSF.  

In the NICMOS images, the most uncertain measures are
found for the compact source SMMJ123632.61+620800.1, and for the merging system
SMMJ123616.15+621513.7, where the NICMOS resolution does not resolve the three
components seen in the ACS observations.  In most of the cases, contaminating neighbours
are present in the image, and we account for these by fitting the surface
brightness profile of these neighbours
together with the target object. As shown by \citet{Hauss:07} the simultaneous fit
allows one to recover structural parameters more reliably than deriving them
using masks.  Likewise, in the case of mergers, the interacting systems are fit
at the same time.  The results of our analysis are presented in Table \ref{tab1}, 
where size, S\'ersic index and axis ratio are shown for the ACS and NICMOS data. 

By comparing sizes and  S\'ersic indices measured in different filters,
we find that the two estimates for the isolated systems agree to within 3$\sigma$. 
The reason for the slight differences in these fits may be ascribed to the morphological 
k-correction,
where different aspects of the stellar populations for these galaxies are observed
at different wavelengths.
This is particularly relevant within actively star forming
galaxies or those with dust, such as the systems we examine in this paper. Note that at the 
median redshift of our sample ($\simeq 2.3$), the NICMOS filter is probing the rest-frame 
optical, while the ACS imaging matches the rest-frame NUV.  Hence the bluer ACS band is 
more sensitive to the clumpy distribution of the star forming regions.  Particularly in the case 
of `disc'-like galaxies, the S\'ersic index measured in the NUV rest-frame is smaller compared to 
the rest-frame optical wavebands (see also \citealt{Rawat:09}). However, in most of the cases 
the NUV and optical rest-frame morphologies agree in their ability to discriminate between 
disc-like ($n<2.5$) and early-type ($n>2.5$) galaxies.

On the other hand, for the merging systems the measurements in the 
NICMOS and ACS bands show significant differences.  One reason for this is the lack of resolution in 
the NICMOS band, which does not always allow us to resolve the merging systems, such as the case of
SMMJ123616.15+621513.7 where three
galaxies are seen in the z-band image, but only two in the H-band image.
The system SMMJ123622.65+621629.7 is a rapidly ongoing merger  which is seen in the rest-frame UV as a 
compact, elongated galaxy with prominent tidal features, while in the NICMOS band the tides are not resolved, 
and the effective radius is much larger. In the case of system SMMJ123635.59+621424.1 the strong disagreement
 is due to the fact that in the z-band we observe a multicomponent system, 
made up of a compact  bulge with a double nucleus, plus faint spiral arms on larger scales. Since the spiral arms 
are too faint to be fit with a single S\'ersic model we have measured the compactness of the bulge. Rather, in the 
NICMOS H-band the multicomponent structure is not resolved, and fits to the system reveal a much more extended profile. 
Hence, merging systems can appear very different at different wavelengths when using images at different resolutions, 
and morphological k-corrections can strongly affect our measurements.

\section{Results}\label{res}

In Figure \ref{acs} we show the ACS images of the sample, classified according to
their morphological properties.  We find that SMGs display a heterogeneous mix of
morphologies.  We divide the sample into three main categories: disc-like objects
(5/12), ongoing mergers (4/12) and compact galaxies (3/12). The first class includes the
majority of the sources (42\%$\pm$18\%), which show quite regular and diffuse morphologies,
light-profiles characteristic of late-type galaxies (i.e. $n<2.5$), and are very faint.
Although we can not rule out that some of them are interacting systems, where the
companion is not detected  due to its faintness at NUV wavelengths \footnote{Indeed two of the disc-like objects, SMMJ123711.98+621325.7 and SMMJ123707.21+621408.1, present multi-component radio counterparts lying at the same redshift \citep{Sw:04, Chapman:05}. Therefore, although they appear isolated in the optical they are likely early stage mergers \citep{Pope:08, Tacconi:08}.}, 
our hypothesis is that
these objects are disc-like galaxies with ongoing SFR. It is important to note that for  
galaxies in this category, where we have NICMOS imaging, the disc-like nature is confirmed 
through the rest-frame optical images. 

The second category, ongoing mergers (33\%$\pm$17\%), are systems where  two or more components are
clearly visible in the ACS band. We include in this class the source 
SMMJ123635.59+621424.1, since a double nucleus is visible, and it is likely to be in the final
stages of a merger. Most of these interacting systems display disc-like structures with a
large range in sizes.  This is also confirmed in NICMOS when data is available. The
final class includes three (25\%$\pm$14\%) very compact, isolated sources.  All of these
objects show a concentrated light-profile, with S\'ersic indices indicative of an early-type morphology
($n>2.5$) and $R_e \lesssim 1$ kpc.  Note that the extremely small sizes measured in the
ACS data (tracing the NUV) indicates that the star formation is extremely concentrated
in the centre of these objects. 
Although the AGN contribution might play a role in shrinking the size measured in the rest-frame UV,  
we find that when NICMOS data are available the near-IR sizes are slightly larger but still 
very small compared to galaxies with similar masses in the nearby universe.

\begin{figure}
\centering
\includegraphics[width=0.9\columnwidth]{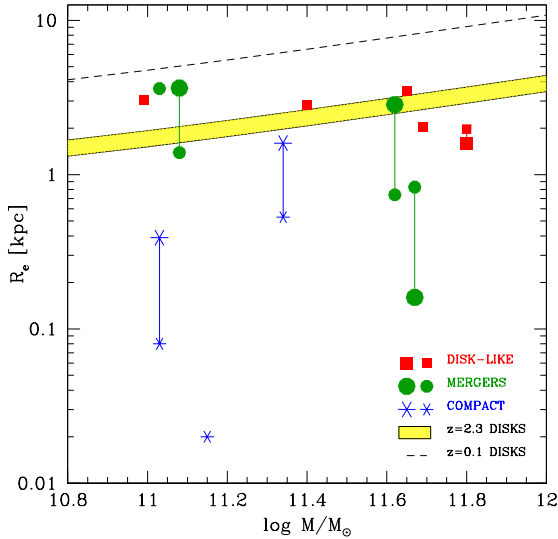}
\caption{Stellar mass-size relation for our sample. Red squares indicate the disc-like
galaxies, green circles are for the 'mergers', and blue asterisks represent the three compact
objects.  Large and small symbols are for NICMOS and ACS data respectively. Overplotted
is the stellar-mass relation for disc galaxies evolved to z=2.3 
(yellow shaded region: upper envelope is the NICMOS-derived relation, lower envelope is the ACS-derived one), following
\citet{Buitrago:08}, and the local relation (black dashed line) from \citet{Shen:03}.  }

\label{RM}
\end{figure}

\begin{figure}
\centering
\includegraphics[width=0.9\columnwidth]{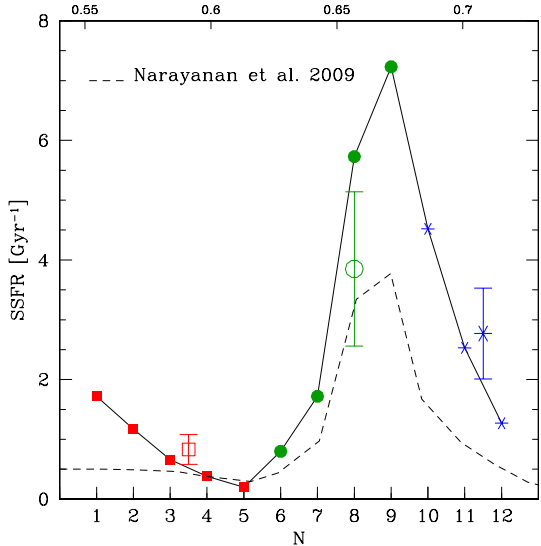}\\
\caption{Specific star formation rate for our 12 SMGs (points, solid line). Different symbols refer 
to different morphological classes as in Figure \ref{RM} (red squares: disc-like galaxies, green circles: 
mergers, blue asterisks: compact galaxies ). 
The open symbols with errorbars indicate the average SSFR for a given class. Average and standard deviation have been computed with a bootstrap resampling.
The x-axis refers to the position of the galaxy in the 
proposed evolutionary sequence shown in Figure \ref{acs}. The dashed line shows the predictions of 
\citet{Narayanan:09} from a merger model for SMGs, with the time-scale shown in the upper axis in units of Gyrs. }
\label{ssfr}
\end{figure}

In Figure \ref{RM} we plot the stellar mass-size relation for our sample galaxies,
divided into the three morphological classes. 
Overplotted is the mass-size relation for disc galaxies from \citet{Buitrago:08} at z=2.3 
(the median redshift of our sample) :

\begin{equation}
R_e(M_*,z=2.3)=\alpha (1+z)^{-\beta}R_e(M_*,z=0.1)\,,
\end{equation}

\noindent with $\alpha=1.1$ and $\beta=0.8$  for NICMOS-derived sizes (upper envelope of the shaded region) and  $\alpha=1.1$ and $\beta=1.0$ 
for the ACS-derived sizes (lower envelope).
The local relation, $R_e(M_*,z=0.1)$, is taken from 
\citet{Shen:03} based on SDSS data. 
Remarkably, the sizes of the disc-like galaxies in our sample match the sizes
found for the general massive galaxy population at this redshift.  
As the relation of \citet{Buitrago:08} does not change significantly for ACS and NICMOS data,
this reinforces the idea that the
ACS sizes of our disc-like objects (around 2-3 kpc) are representative of the rest-frame
optical sizes of these galaxies.

We note that some of the galaxies in our sample were already studied in previous works 
\citep{Chapman:03, Cons:03, Smail:04, Almaini:05, Pope:05} 
using both ground-based and HST data. They found a higher merger fraction ($\simeq 50-60 \%$) 
with respect to ours ($\simeq 30\%$), partly due to the method adopted for the classification 
of mergers. While the above authors use schemes based on the asymmetry parameter,
in our analysis we only considered ongoing mergers those systems where two or more distinct 
components are resolved in the ACS image. As mentioned above, 
it is also possible that dust is hiding the interacting structure, and we are missing a fraction 
of actual mergers, hence our measured merger fraction has to be considered a lower limit.

\section{Discussion}\label{disc}

Motivated by high-resolution hydrodynamical simulations \citep{Dekel:06, Cox:08,
Hopk:09, Narayanan:09}, we argue that  the three morphological classes outlined above
could represent an evolutionary sequence, where the disc-like class represents a
pre-merger phase followed by the major merger event, while  the compact sources can be
interpreted as the end-stages of the merger, caught during or just after the
coalescence. Numerical simulations naturally explain how the tidal forces involved in
the interactions remove angular momentum from the systems, allowing the gas to fuel
towards the centre. At this stage, the gas is compressed into a very small volume, leading
to surface densities of the order $\simeq 10^{5} M_{\odot} pc^{-2}$ \citep{Hopk:09b}, close to that of
molecular clouds.  According to the Kennicutt law, the SFR in such conditions is
extremely enhanced.  Since the dynamical time-scale that drives the collapse is similar
to the star formation rate, the system rapidly exhausts its gas while it contracts
\citep{MH:96, Hopk:08}. Thus, it is not surprising that we observe such compact galaxies
with high SFR. The models, indeed, predict a spheroidal-like morphology at the time of
coalescence or just after, since the coalescence generally completes at about the same time
as the gas first reaches the centre \citep{Cox:08p}.

Moreover, our findings that a fraction of the SMGs have a compact morphology agrees with 
measurements of the gas distribution from CO maps for many galaxies \citep{Tacconi:06, Tacconi:08}. 
Therefore, these compact, highly star-forming systems are likely to be in the final phase of 
the merger, and are the transition link between starbursts and compact galaxies.

A peculiar case is represented by the system SMMJ123635.59+621424.1.
We classify it as a merger event, as it appears as a double-nucleus system. However, its effective radius 
and S\'ersic index are characteristic of those for compact galaxies.  It is likely that we are measuring the 
compactness of the bulge component, given the faintness of the outer light. This systems also shows faint signs 
of potential spiral structure over a scale of 5 kpc. The SSFR of 
this system is the highest in our sample, compatible with being near the peak in star formation during 
the coalescence phase. It is worth to note that this object has the highest infrared luminosity 
$L=10^{13.01}L_{\odot}$ (from \citealt{Mich:09}) of our sample, hence it can be considered a HLIRG,
supporting the picture where HLIRGs are galaxies in their maximal star formation periods triggered by
interactions.

In order to build an ``illustrative'' evolutionary sequence, we can order galaxies inside a given class 
according to the values of their SSFR. In Figure \ref{ssfr} we show as a dashed line, the evolution of the SSFR 
in a simulated galaxy merger, taken from \citet{Narayanan:09}.
The simulation illustrates the evolution of the SFR in a major merger (with mass ratio 1:1) for a $\sim 2 
\times 10^{13} M_{\odot}$ dark matter halo. 
 We have computed the SSFR taking the SFR of their Figure 1 and dividing it by the final stellar mass 
($8\times10^{11}M_{\odot}$, which is roughly the maximum stellar masses of our sample). 
 The trend in SSFR shows a modest star formation rate in the pre-merger phase, in agreement with 
the value of our diffuse galaxies, followed by a steep rise during the starburst/merger, a peak in the 
coalescence phase, and then a rapid decline. The trend depicted by the simulations is matched by our data 
if we assume that the SSFR increases in the merging phase, and then declines for the compact galaxies. 
The peak is reached  during the coalescence, as it is illustrated by the source SMMJ123635.59+621424.1. 
In simulations the diffuse/isolated systems are not fuelled by a new gas reservoir, as in the case for 
merging systems, and they simply exhaust the cold gas available.  Hence their SSFR declines with time (see \citealt{Cox:08}).
Therefore, in the same plot, we have indicated by the coloured points the SSFR of our sample of SMGs, with 
the number in the x-axis indicating the position in the evolutionary sequence (in Figure \ref{acs} galaxies 
are ordered according to this sequence). 
Note that we adopt the SFR derived from the IR luminosity from \citet{Mich:09}, using the \citet{Kennicutt:98} relation
to computed the star formation rate.

The fact that we can put our galaxy sample in a SSFR-sequence that well matches that found in hydrodynamical 
simulations, strengthens the conclusion 
that the (morphological) evolutionary sequence described above, where large diffuse systems transform 
themselves in compact remnants passing through  a major merger event, is likely the formation mechanism
for these galaxies into compact systems seen at slightly lower redshifts.
Another piece of evidence supporting this theoretical merging scenario is the size of the
progenitor discs we have found. Theoretically these discs are expected  to have
effective radius of 2-3 kpc \citep{Dekel:06, Hopk:09}. This is in fact what our
observations show. 
Given the lack of NICMOS imaging for most of the sources and the  insufficient resolution in the near-IR,
we were forced to use ACS imaging to constrain our morphological sequence.
To better test such picture high-sensitivity imaging in near-IR would be required, and this could be achieved only when  WFC3 imaging will be available.

Summarising, our data can not reject the evolutionary picture
depicted by theoretical models in which the precursors of the superdense galaxies are massive, gas-rich discs at 
$z \sim 2-3$, which evolve into compact remnants through dissipative major mergers. 
Moreover, a cold gas accretion driven scenario for the formation of the compact massive galaxies, as the one proposed by \citet{Dekel:09}
perhaps can not be easily supported by our data, since in this case we would likely not observe such a large diversity of size
and structure for these progenitors as we observe here.

\section*{Acknowledgements}

We thank the referee for helpful suggestions which improved the paper. 
We are grateful to Philip F. Hopkins for insightful comments in relation to some
theoretical issues. 
We  also thank Kar\'in Men\'endez-Delmestre for her suggestions, and
Michele Cirasuolo, Loretta Dunne and Nathan Bourne for useful 
discussions.

\clearpage
\begin{landscape}
\begin{table}
\begin{minipage}{200mm}
\caption{Properties of SMGs }
\begin{tabular}{lcccccccc|ccccc}
\hline \hline
&&&&&\multicolumn{4}{c}{{\bf ACS}} & \multicolumn{4}{c}{{\bf NICMOS}}\\
\cline{6-9} \cline{10-13}
Galaxy \footnote{Galaxy ID, objects without ID are the subcomponents in the merging systems.} & z \footnote{Spectroscopic redshift from \citet{Chapman:05}.} & $\log \frac{M}{M_{\odot}}$ \footnote{Stellar mass from \citet{Mich:09} converted to a Kroupa IMF.} & SSFR  \footnote{Specific star formation rate from \citet{Mich:09}.} & $R_{AB}$ \footnote{Magnitudes in the R-band (AB system) from \citet{Chapman:05}} & $R_e$ & $R_e$ & n & $b/a$ \hspace{11mm}\rule[-2mm]{0.1mm}{6mm} &  $R_e$ & $R_e$ & n & $b/a$ & Class \footnote{Morphological class: D stays for disc-like objects, M for ongoing mergers and C for compact galaxies (see Sect. 4).}\\
&&& [Gyr$^{-1}$] && arcsec & [kpc] &&  \hspace{16.45mm}\rule[-2mm]{0.1mm}{6mm} & arcsec & [kpc] &&\\
\cline{6-9} \cline{10-13}

SMMJ123711.98+621325.7 & 1.992 & 10.99 & 1.72 & 25.8 & 0.36 & 3.03$\pm$0.03 & 0.69$\pm$0.03 & 0.79$\pm$0.00 \footnote{Measurements with errors equal to 0.00 mean that the parameters have been fixed in the fitting procedure.}  \hspace{-0.1mm}\rule[-2mm]{0.1mm}{6mm}
&  &  &  & & D\\
SMMJ123606.85+621021.4 & 2.509 & 11.69 & 1.17 & 25.3 & 0.25 & 2.03$\pm$0.04 & 2.38$\pm$0.00 & 0.70$\pm$0.00 \hspace{2.5mm}\rule[-2mm]{0.1mm}{6mm}
&  &  &  & & D\\
SMMJ123618.33+621550.5 & 1.865 & 11.40 & 0.66 & 25.9 & 0.34 &2.84$\pm$0.12 & 2.19$\pm$0.14 & 0.66$\pm$0.12 \hspace{2.5mm}\rule[-2mm]{0.1mm}{6mm}
&  &  &  & & D\\
SMMJ123707.21+621408.1 & 2.484 & 11.65 & 0.38 & 26.0 & 0.43 & 3.46$\pm$0.39 & 2.54$\pm$0.73 & 0.43$\pm$0.11 \hspace{2.5mm}\rule[-2mm]{0.1mm}{6mm}
&  &  &  & & D\\
SMMJ123712.05+621212.3 & 2.914 & 11.80 & 0.20 & 25.5 & 0.25 & 1.97$\pm$0.04 & 0.98$\pm$0.02 & 0.67$\pm$0.00 \hspace{2.5mm}\rule[-2mm]{0.1mm}{6mm}
& 0.21 & 1.60$\pm$0.11 & 1.51$\pm$0.36 & 0.72$\pm$0.09 & D \\
SMMJ123616.15+621513.7 & 2.578 & 11.67 & 0.80 & 25.7 & 0.10 & 0.83$\pm$0.05 & 2.64$\pm$0.76 & 0.74$\pm$0.07 \hspace{2.5mm}\rule[-2mm]{0.1mm}{6mm}
& 0.02 & 0.16$\pm$0.12 & 3.91$\pm$2.69 & 0.47$\pm$0.43 & M\\
&&&&& 0.07 & 0.53$\pm$0.04 & 0.85$\pm$0.30 & 0.53$\pm$0.05 \hspace{2.5mm}\rule[-2mm]{0.1mm}{6mm}
& &  & & \\
&&&&& 0.11 & 0.86$\pm$0.07 & 0.69$\pm$0.13 & 0.66$\pm$0.06  \hspace{2.5mm}\rule[-2mm]{0.1mm}{6mm}
& 0.28 & 2.23$\pm$0.17 & 1.00$\pm$0.00 & 0.97$\pm$0.05 \\
SMMJ123622.65+621629.7 & 2.466 & 11.62 & 1.72 & 25.4 & 0.09 & 0.74$\pm$0.04 & 1.50$\pm$0.02 & 0.12$\pm$0.01 \hspace{2.5mm}\rule[-2mm]{0.1mm}{6mm}
& 0.35 & 2.84$\pm$0.07& 0.37$\pm$0.02 & 0.41$\pm$0.03 & M\\
&&&&& 0.58 & 4.66$\pm$0.43 & 1.96$\pm$0.17 & 0.10$\pm$0.00 \hspace{2.5mm}\rule[-2mm]{0.1mm}{6mm}
& 0.54 & 4.38$\pm$0.73& 2.23$\pm$0.04 & 0.35$\pm$0.04 \\
SMMJ123553.26+621337.7 & 2.098 & 11.03 & 5.73 & 24.7 & 0.43 & 3.61$\pm$0.15 & 2.75$\pm$0.12 & 0.28$\pm$0.00  \hspace{2.5mm}\rule[-2mm]{0.1mm}{6mm}
&  &  &  & &M\\
&&&&& 0.14 & 1.12$\pm$0.06 & 0.51$\pm$0.04 & 0.22$\pm$0.00 \hspace{2.5mm}\rule[-2mm]{0.1mm}{6mm}
&  &  & &\\
SMMJ123635.59+621424.1 & 2.005 & 11.08 & 7.23 & 24.2 & 0.17 & 1.39$\pm$0.28 & 0.87$\pm$0.00 & 0.78$\pm$0.00 \hspace{2.5mm}\rule[-2mm]{0.1mm}{6mm}
& 0.44 & 3.64$\pm$0.12 & 1.97$\pm$0.08 & 0.81$\pm$0.04 & M \\
SMMJ123632.61+620800.1 & 1.993 & 11.03 & 4.52 & 23.6 & 0.01 & 0.08$\pm$0.04 & 4.65$\pm$1.22 & 1.00$\pm$0.00 \hspace{2.5mm}\rule[-2mm]{0.1mm}{6mm}
 & 0.05 & 0.39$\pm$0.73 & 2.78$\pm$3.33 & 0.45$\pm$0.42 & C\\
SMMJ123600.15+621047.2 & 1.994 & 11.34 & 2.53 & 25.1 & 0.06 & 0.53$\pm$0.10 & 3.57$\pm$0.66 & 1.00$\pm$0.00 \hspace{2.5mm}\rule[-2mm]{0.1mm}{6mm}
& 0.19 & 1.60$\pm$0.18 & 0.63$\pm$0.28 & 1.00$\pm$0.00 & C\\
SMMJ123606.72+621550.7 & 2.416 & 11.15 & 1.27 & 23.6 & 0.002 & 0.02$\pm$0.02 & 4.47$\pm$3.12 & 1.00$\pm$0.00 \hspace{2.5mm}\rule[-2mm]{0.1mm}{6mm}
&  &  &  & &C\\
\hline \hline

\label{tab1}
\end{tabular}
\end{minipage}
\end{table} 
\end{landscape}

\label{lastpage}

\end{document}